\documentclass[journal]{IEEEtran}
\usepackage{latexsym}

\usepackage{enumerate}
\usepackage{graphicx}
\usepackage{subfigure}
\usepackage{multirow}

\usepackage{framed,multirow}
\usepackage{times}
\usepackage{graphicx} 
\usepackage{subfigure}
\usepackage{amssymb}
\usepackage{latexsym}

\usepackage{algorithm}
\usepackage{algorithmic}

\usepackage{hyperref}

\usepackage{url}
\usepackage{xcolor}
\usepackage{amsmath}
\usepackage{amsxtra}
\usepackage{braket}
\usepackage{array}
\usepackage{color}
\usepackage{url}
\usepackage{placeins}
\usepackage{threeparttable}

\usepackage{times}
\usepackage{balance}
\usepackage{enumerate}
\usepackage{multirow}
\usepackage{pifont}
\usepackage{marvosym}
\usepackage{ifsym}

\begin{document}
\title{A Hierarchical Transitive-Aligned Graph Kernel for Un-attributed Graphs}

\author{Lu~Bai${}^{1}$,~\IEEEmembership{}Lixin~Cui${}^{1*}$,~\IEEEmembership{}and~Edwin~R.~Hancock${}^{3}$,~\IEEEmembership{~IEEE~Fellow}

\thanks{Lu Bai, Lixin Cui${}^{*}$ (Corresponding Author: cuilixin@cufe.edu.cn) are with ${}^{1}$Central University of Finance and Economics, Beijing, China. Edwin R. Hancock is with ${}^{3}$University of York, York, UK.}
}
\markboth{Journal of \LaTeX\ Class Files,~Vol.~6, No.~1, January~2007}%
{Shell \MakeLowercase{\textit{et al.}}: Bare Demo of IEEEtran.cls
for Journals}
\maketitle

\begin{abstract}
In this paper, we develop a new graph kernel, namely the Hierarchical Transitive-Aligned kernel, by transitively aligning the vertices between graphs through a family of hierarchical prototype graphs. Comparing to most existing state-of-the-art graph kernels, the proposed kernel has three theoretical advantages. First, it incorporates the locational correspondence information between graphs into the kernel computation, and thus overcomes the shortcoming of ignoring structural correspondences arising in most R-convolution kernels. Second, it guarantees the transitivity between the correspondence information that is not available for most existing matching kernels. Third, it incorporates the information of all graphs under comparisons into the kernel computation process, and thus encapsulates richer characteristics. By transductively training the C-SVM classifier, experimental evaluations demonstrate the effectiveness of the new transitive-aligned kernel. The proposed kernel can outperform state-of-the-art graph kernels on standard graph-based datasets in terms of the classification accuracy.
\end{abstract}


\IEEEpeerreviewmaketitle

\section{Introduction} \label{s1}

Graph-based representations are powerful tools to represent structure data that is described with pairwise relationships between components. The main challenge arising in analyzing the graph-based data is how to learn effective numeric features of the discrete graph structures. One way to achieve this is to employ graph kernels, that can characterize graph structures in a high dimensional space and thus better preserve the structure information~\cite{haussler99convolution}.

\subsection{Related Works} \label{s1.1}

In machine learning, a graph kernel is defined in terms of a similarity measure between graph structures. One of the most successful and widely used approach to defining kernels between a pair of graphs is to decompose the graphs into substructures and to compare/count pairs of specific isomorphic substructures~\cite{haussler99convolution}. Specifically, any graph decomposition can be used to define a kernel, e.g., the graph kernel based on comparing all pairs of decomposed a) walks, b) paths and c) restricted subgraph or subtree structures. With this scenario, Kashima et al.~\cite{DBLP:conf/icml/KashimaTI03}
have proposed a Random Walk Kernel by comparing pairs of isomorphic
random walks in a pair of graphs. Borgwardt et al.~\cite{DBLP:conf/icdm/BorgwardtK05} have proposed a Shortest Path
Kernel by counting the numbers of pairwise shortest paths having the
same length in a pair of graphs. Costa and Grave~\cite{DBLP:conf/icml/CostaG10} have defined a Neighborhood Subgraph
Pairwise Distance Kernel by counting the number of pairwise
isomorphic neighborhood subgraphs. Gaidon et al.~\cite{DBLP:conf/bmvc/GaidonHS11} have developed a Subtree Kernel for comparing videos, by considering complex actions as decomposed spatio-temporal parts and building corresponding binary trees. The resulting kernel is computed by counting the number of isomorphic subtree patterns. Other alternative graph kernels that are specifically based on the R-convolution framework also include a) the Segmentation Graph Kernel~\cite{DBLP:conf/cvpr/HarchaouiB07}, b) the Pyramid Quantized Weisfeiler-Lehman Kernel~\cite{DBLP:journals/ijon/GkirtzouB16}, c) the Subgraph Matching Kernel~\cite{DBLP:conf/icml/KriegeM12}, d) the Quantum-inspired Jensen-Shannon Kernel~\cite{DBLP:journals/tcyb/Bai20}, etc.

One major drawback arising in most existing R-convolution kernels is that they neglect the relative locational information between substructures. Specifically, the R-convolution kernels usually tend to add an unit value when a pair of similar substructures are identified. However, these kernels cannot identify whether these similar substructures are correctly aligned with the overall graph structures, i.e., they do not check if the topological arrangement of the substructures is globally correct. For an instance of a protein matching problem, we may have similar substructures from different parts of the overall structure. R-convolution kernels will count these as being matching substructures, despite the fact that they are not correctly aligned. To overcome this drawback, Bai et al.~\cite{DBLP:conf/ijcai/BaiZW0H15,DBLP:conf/icml/Bai0ZH15} have developed a family of novel vertex-based matching kernels by aligning depth-based representations of vertices~\cite{DBLP:journals/pr/BaiH14}. All these matching kernels can be seen as aligned subgraph or subtree kernels that incorporate explicit structural correspondences, and thus address the drawback of neglecting relative locations between substructures arising in the R-convolution kernels. Unfortunately, these matching kernels are not positive definite in general. This is because the alignment steps for these kernels are not transitive. In other words, if $\sigma$ is the vertex-alignment between graph $A$ and graph $B$, and $\pi$ is the alignment between graph $B$ and graph $C$, in general we cannot guarantee that the alignment between graph $A$ and graph $C$ is $\pi \circ \sigma$. On the other hand, Fr{\"{o}}hlich et al.~\cite{DBLP:conf/icml/FrohlichWSZ05} have demonstrated that the transitive alignment step is necessary to guarantee the positive definiteness of the vertex/edge based matching kernels. Furthermore, either the R-convolution kernels or the matching kernels only capture graph characteristics for each pair of graphs, and thus ignore the information over other graphs. As a summary, developing effective graph kernels still remains challenges.

\begin{figure*}
\centering
\subfigure{\includegraphics[width=1\linewidth]{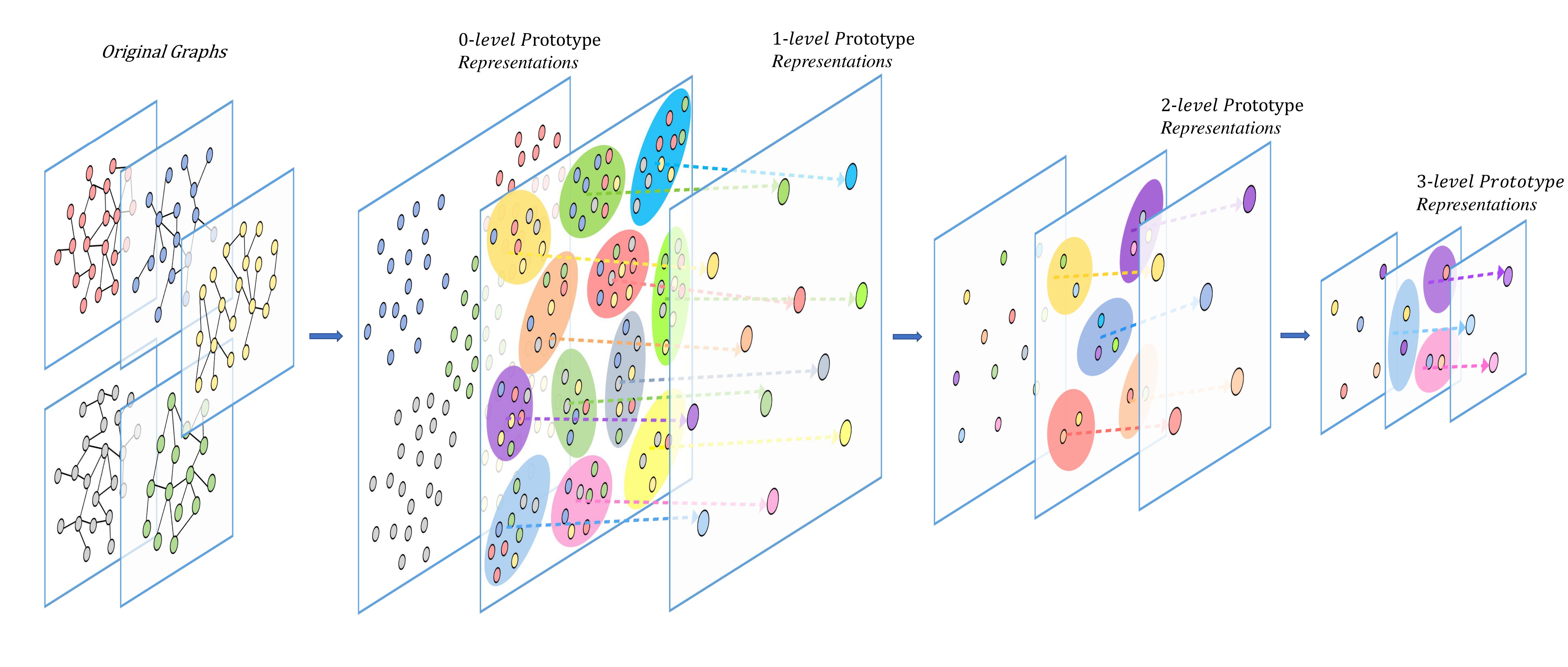}}
\vspace{-30pt}
\caption{\small{The framework of constructing the hierarchical prototype representations. For a set of five original graphs, we commence by employing their $k$-dimensional vectorial representations as the $0$-level prototype representations (in this instance, $k=2$). By employing the $\kappa$-means method, we hierarchically identify a set of centroid points (i.e., means) as the $h$-level prototype representations over the set of the last $h$-level prototype representations, where $h$ varies from 1 to 3.}} \label{HPR_framework}
\vspace{-10pt}
\end{figure*}

\subsection{Contributions} \label{s1.1}

The aim of this work is to address the aforementioned shortcomings of existing graph kernels, by developing a new Hierarchical Transitive-Aligned Kernel (HTAK) for un-attributed graphs. The key innovation of the proposed kernel is that of transitively aligning vertices between pairs of graphs, through a family of hierarchical prototype representations. That is, given three vertices $v$, $w$ and $x$ from three different sample graphs, if $v$ and $x$ are aligned, and $w$ and $x$ are aligned, the proposed kernel can guarantee that $v$ and $w$ are also aligned. As a result, the proposed kernel can theoretically guarantee the positive definiteness. Specifically, the main contributions of this work are threefold.


\textbf{First}, we propose a framework to compute a family of $H$-hierarchical prototype representations that encapsulate the dominant characteristics of the vectorial vertex representations over a set of graphs $\mathbf{G}$. This is achieved by hierarchically performing the $\kappa$-means clustering method to identify a preassigned number of cluster centroid as the $h$-level prototype representations through the last $h-1$-level prototype representations, where the $0$-level representations correspond to the original vectorial vertex representations of all graphs. This in turn generate a family of $H$-hierarchical prototype representations, when we vary $h$ from $1$ to $H$ (i.e., $1\leq h\leq H$). We show that the new hierarchical prototype representations not only reflect the general structural information over all graphs, but also represent a reliable pyramid of vertices over all graphs at different levels.


\textbf{Second}, with the family of $H$-hierarchical prototype representations to hand, we develop a graph matching method by hierarchically aligning the vertices of each graph to its different $h$-level prototype representations. The resulting HTAK kernel is defined by counting the number of aligned vertex pairs. We show that the proposed kernel not only overcomes the shortcoming of ignoring correspondence information between isomorphic substructures that arises in most existing R-convolution kernels, but also guarantees the transitivity between the correspondence information. As a result, the proposed kernel guarantees positive definite that is not available in existing alignment kernels~\cite{DBLP:conf/ijcai/BaiZW0H15,DBLP:conf/icml/Bai0ZH15}. Furthermore, unlike most existing graph kernels, the proposed kernel incorporates the information of all graphs under comparisons into the kernel computation process, and thus encapsulates richer characteristics.

\textbf{Third}, by transductively training the C-SVM classifier associated with the proposed HTAK kernel, we empirically demonstrate the effectiveness of the new kernel appraoch. The proposed kernel can outperform state-of-the-art graph kernels as well as graph neural network models on standard graph datasets in terms of the classification accuracy.

The remainder of this paper is organized as follows. Section~\ref{s2} introduces the framework of computing the hierarchical prototype representations. Section~\ref{s3} gives the definition of the new kernel, Section~\ref{s4} provides experimental evaluations and Section~\ref{s5} concludes the work.



\section{Hierarchical Prototype Representations}\label{s2}

In this section, we propose a framework to compute a family of $H$-hierarchical prototype representations that encapsulate the dominant characteristics over all vectorial vertex representations in a set of graphs $\mathbf{G}$. An instance of the proposed framework to compute the hierarchical prototype representations is shown in Fig.\ref{HPR_framework}. Specifically, let $$\mathbf{{R}}^k =\{\mathrm{R}_1^k,\mathrm{R}_2^k,\ldots,\mathrm{R}_i^k,\ldots,\mathrm{R}_N^k\}$$ denote the $k$-dimensional vectorial representations of $N$ vertices over all graphs in $\mathbf{G}$. We first adopt $\mathbf{{R}}^k$ as the set of $0$-level prototype representations $\mathbf{{PR}}^{0,k}$, i.e.,
\begin{equation}
\mathbf{{PR}}^{0,k}=\{\mathrm{PR}_1^{0,k},\mathrm{PR}_2^{0,k},\ldots,\mathrm{PR}_i^{0,k},\ldots,\mathrm{PR}_{N_0}^{0,k}\},
\end{equation}
where all the $0s$ indicate the current value of the parameter $h$, $N=N_0$, and each $i$-th element $\mathrm{PR}_i^{0,k}$ corresponds to $\mathrm{R}_i^{k}$. To compute the set of the higher $h$-level (i.e., $1\leq h\leq H$) prototype representations $\mathbf{{PR}}^{h,k}$, we employ $\kappa$-means~\cite{witten2011data} to localize $N_h$ centroid points over the set of the last lower $h-1$-level prototype representations $\mathbf{{PR}}^{h-1,k}$, by minimizing the objective function
\begin{equation}
\arg\min_{\Omega}  \sum_{j=1}^{N_h} \sum_{\mathrm{PR}_i^{h-1,k} \in c_j} \|\mathrm{PR}_i^{h-1,k}- \mu_j^k \|^2_2,\label{kmeans}
\end{equation}
where $\Omega=(c_1,c_2,\ldots,c_j,\ldots,c_{N_h})$ represents $N_h$ clusters over the set of $h-1$-level prototype representations $\mathbf{{PR}}^{h-1,k}$, and $\mu_j^k$ is the mean of the prototype representations belonging to the $j$-th cluster $c_j$. We employ the $N_h$ means $\{\mu_j^1,\mu_j^2, \ldots,\mu_j^k,\ldots,\mu_{N_h}^k \}$ as the set of $h$-level prototype representations $\mathbf{{PR}}^{h,k}$, i.e.,
\begin{equation}
\mathbf{{PR}}^{h,k}=\{\mathrm{PR}_1^{h,k},\mathrm{PR}_2^{h,k},\ldots,\mathrm{PR}_j^{h,k},\ldots,\mathrm{PR}_{N_h}^{h,k}\},
\end{equation}
where each $j$-th element $\mathrm{PR}_j^{h,k}$ corresponds to $\mu_j^k$, and $N_h$ corresponds to the number of the $h$-layer prototype representations in $\mathbf{{PR}}^{h,k}$.

Since the value of $N_h$ (i.e., $|\mathbf{{PR}}^{h,k}|$) is usually much lower than that of $N_{h-1}$ (i.e., $|\mathbf{{PR}}^{h-1,k}|$), the initialized set of $0$-level prototype representations $\mathbf{{PR}}^{0,k}$ correspond to the original vectorial representations of the vertices over all graphs in $\mathbf{G}$, and the set of $h$-level prototype representations $\mathbf{{PR}}^{h,k}$ are computed through the objective function of $\kappa$-means (i.e., Eq.(\ref{kmeans})) that can gradually minimize the inner-vertex-cluster sum of squares over the set of the last $h-1$-level prototype representations $\mathbf{{PR}}^{h-1,k}$. When we vary the parameter $h$ from $1$ to $H$, this procedure naturally forms a family of $H$-hierarchical prototype representations as
\begin{equation}
\mathbb{{PR}}^{H,k}=\{\mathbf{{PR}}^{1,k},\mathbf{{PR}}^{2,k}, \ldots,\mathbf{{PR}}^{h,k},\ldots, \mathbf{{PR}}^{H,k}\},
\end{equation}
where each $\mathbf{{PR}}^{h,k}$ is the set of $h$-level prototype representations, and $\mathbb{{PR}}^{H,k}$ represents a reliable pyramid of the original vertex representations over all graphs at different levels (i.e., the prototype representations of different $h$-levels).

\begin{figure*}
\centering
\subfigure{\includegraphics[width=1\linewidth]{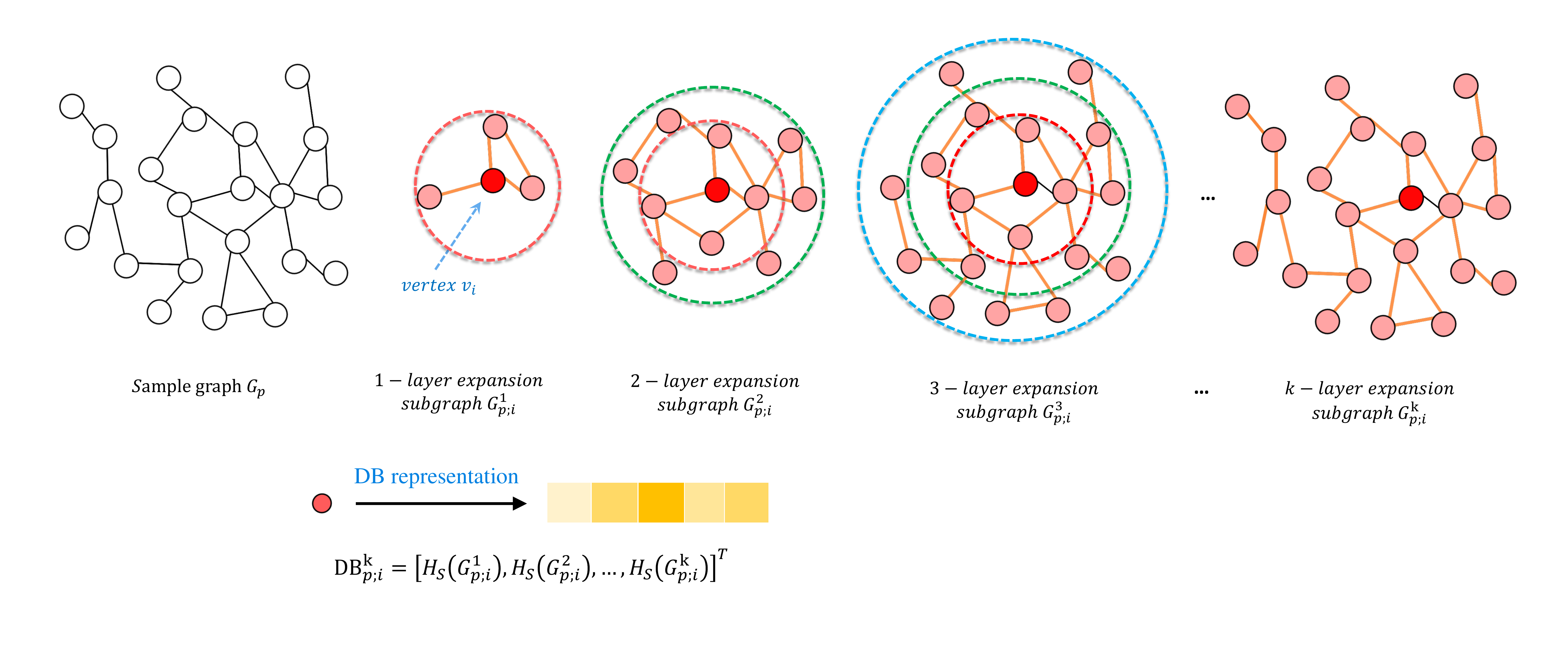}}
\vspace{-45pt}
\caption{\small{The process of computing the depth-based representation rooted at each vertex.}} \label{DBR}
\vspace{-10pt}
\end{figure*}

Note that, to compute the family of $H$-hierarchical prototype representations, in this work we employ the $k$-dimensional depth-based (DB) representations as the original $k$-dimensional vectorial vertex representations $\mathbf{{R}}^k$ (i.e., $\mathrm{PR}_i^{0,k}$) to compute the different sets of $h$-level prototype representations $\mathbf{{PR}}^{h,k}$. Certainly, computing the vertex representations is an open problem, on can also utilize any other approach to compute the initialized vectorial vertex representations~\cite{DBLP:journals/pami/WilsonHL05,DBLP:journals/pr/XiaoHW09}. Specifically, in this work, the specified DB representation of each vertex is defined by measuring the entropies on a family of $\widetilde{k}$-layer expansion subgraphs rooted at the vertex~\cite{DBLP:journals/pr/BaiH14}, where $\widetilde{k}$ varies from $1$ to $k$. Since each $\widetilde{k}$-layer expansion subgraph completely contains the whole topological structure of the $\widetilde{k}-1$-layer expansion subgraph, it is shown that such a $k$-dimensional DB representation encapsulates rich entropic content flow from each local vertex to the global graph structure, as a function of depth. Fig.\ref{DBR} exhibits the detailed process of computing the DB representation. Specifically, for each sample graph $G_p(V_p, E_p) \in \mathbf{G}$ indicated by the black color and its $i$-th vertex $v_i$ indicated by the red color in Fig.\ref{DBR}, we commence by computing the $1$-layer neighborhood set $\mathcal{N}_{i}^1$ as $$\mathcal{N}_{i}^1=\{v_j \in V_p\ |\ s(v_i,v_j)\leq 1\},$$ where $s(v_i,v_j)$ is the shortest path between the $i$-th vertex $v_i$ and the $j$-th vertex $v_j$. The resulting $1$-layer expansion subgraph $\mathcal{G}_{p;i}^1$ is defined as the substructure preserving the vertices in $\mathcal{N}_{i}^1$ as well as the edges between them from the original global graph $G_p$, i.e., the substructures surrounded by the red broken line in Fig.\ref{DBR}. Similarly, we also construct the $2$-layer and $3$-layer expansion subgraphs surrounded by the green and blue broken lines respectively in Fig.\ref{DBR}. By parity of reasoning, we generate a family of $\widetilde{k}$-layer expansion subgraphs rooted at $v_i$ ($1\leq \widetilde{k} \leq k$). Note that, if $k$ is greater than the longest shortest path rooted from $v_i$ to the remaining vertices of $G_p$, the $k$-layer expansion subgraph $\mathcal{G}_{p;i}^k$ is the global structure of $G_p$. The resulting $k$-dimensional DB representation rooted at $v_i$ is  $${\mathrm{DB}}^k_{p;i}=\{H_S(\mathcal{G}_{p;i}^1),\cdots,H_S(\mathcal{G}_{p;i}^{2}),\cdots,H_S(\mathcal{G}_{p;i}^{k})]^T,$$ where $H_S(\cdot)$ is the Shannon entropy of a (sub)graph associated with the steady state random walk~\cite{DBLP:conf/icml/Bai0ZH15}.

\section{Hierarchical Transitive-Aligned Kernels}\label{s3}
In this section, we propose a novel Hierarchical Transitive-Aligned Kernel (HTAK) for un-attributed graphs. We commence by introducing a new hierarchical transitive vertex matching method, through the family of $H$-hierarchical prototype representations. Moreover, we develop the HTAK kernel based on the new vertex matching method.


\subsection{Hierarchical Transitive Vertex Matching Methods}\label{GraphMatching}
In this subsection, we develop a new hierarchical transitive vertex matching method, by hierarchically aligning the vertices of each graph to each set of $h$-level prototype representations from the family of $H$-hierarchical prototype representations defined in Section~\ref{s2}. For a set of $T$ graphs $\mathbf{G}=\{G_1,\ldots,G_{T}\}$, we commence by computing the family of $H$-hierarchical prototype representations over the $k$-dimensional vectorial vertex representations of all $T$ graphs as $\mathbb{{PR}}^{H,k}=\{\mathbf{{PR}}^{1,k},\ldots,\mathbf{{PR}}^{h,k},\ldots, \mathbf{{PR}}^{H,k}\}$. To establish the correspondence information between the graph vertices, we align the vectorial vertex representations of a sample graph $G_p(V_p,E_p)\in \mathbf{G}$ to each set of $h$-level prototype representations $\mathbf{{PR}}^{h,k}=\{\mathrm{PR}_1^{h,k},\ldots,\mathrm{PR}_n^{h,k},\ldots,\mathrm{PR}_{N_h}^{h,k}\}$. The alignment process is similar to that introduced in~\cite{DBLP:conf/icml/Bai0ZH15} for point matching in a pattern space. Specifically, we compute a $h$-level affinity matrix in terms of the Euclidean distances between the two sets of points as
\begin{equation}
R^{h,k}_{p}(i,n)=\|{R}_{p;i}^k-      \mathrm{{PR}}^{h,k}_n      \|_2,\label{AffinityM}
\end{equation}
where $R^{h,k}_{p}$ is a ${|V_p|}\times {N_h}$ matrix, and each element $R^{h,k}_{p}(i,n)$ represents the distance between the $k$-dimensional vectrial representation ${R}_{p;i}^k$ of $v_i \in V_p$ and the $n$-th $h$-level prototype representation $\mathrm{{PR}}^{h,k}_n\in \mathbf{{PR}}^{h,k}$. For the affinity matrix $R^{h,k}_{p}$, the rows index the vertices of $G_p$, and the columns index the $h$-level prototype representations in $\mathbf{{PR}}^{h,k}$. If $R^{h,k}_{p}(i,n)$ is the smallest element in column $n$, we say that the $k$-dimensional vectorial representation of $v_{i}$ is aligned to the $n$-th $h$-level prototype representation $\mathrm{{PR}}^{h,k}_n\in \mathbf{{PR}}^{h,k}$.

Similarly, for each other sample graph $G_q(V_q,E_q)\in \mathbf{G}$, we also align its $k$-dimensional vectorial representation ${R}_{q;j}^{k}$ of each vertex $v_j\in V_q$ to each set of $h$-level prototype representations $\mathbf{{PR}}^{h,k}$. We compute the element $R^{h,k}_{q}(j,n)$ for the corresponding affinity matrix $R^{h,k}_{q}$ as
\begin{equation}
R^{h,k}_{q}(j,n)=\|{R}_{q;j}^k-      \mathrm{{PR}}^{h,k}_n      \|_2.\label{AffinityM2}
\end{equation}

\noindent\textbf{Definition (Vertex matching between a pair of graphs):} For the pair of graphs $G_p$ and $G_q$ of $\mathbf{G}$, if $R^{h,k}_{p}(i,n)$ and $R^{h,k}_{q}(j,n)$ are both the smallest elements in columns $n$ of $R^{h,k}_{p}$ and $R^{h,k}_{q}$ respectively, we say that the vertex $v_{i}$ of $G_p$ and the vertex $v_{j}$ of $G_q$ are aligned, i.e., there is an one-to-one correspondence between $v_{i}$ and $v_{j}$. More formally, let the $h$-level correspondence matrix $M^{(h,k)}_{p}\in \{0,1\}^{|V_p|\times N_h}$ record the state of alignments for $R^{h,k}_{p}$, and
\begin{equation}
M^{h,k}_{p}(i,n)=\left\{
\begin{array}{cl}
1   & \mathrm{if} \  R^{h,k}_{p} (i,n) \ \mathrm{is \ the \ smallest } \\
    & \mathrm{element \ in \ row} \ n, \ \mathrm{and} \  |{\mathcal{S}}_{{i}}^k|\neq 0;\\
0   & \mathrm{otherwise}.
\end{array} \right.
\label{CoMatrix1}
\end{equation}
Note that, $\mathcal{S}_{{i}}^k$ indicates the set of vertices having the shortest path of length $k$ to $v_i$, and the condition $|\mathcal{S}_{{i}}^k|\neq 0$ guarantees that the $k$-layer expansion subgraph rooted at $v_{i}$ does not surpass the global structure of $G_p$ (i.e., the $k$-dimensional DB representation of $v_i$ exists). Similarly, the $h$-level correspondence matrix $M^{(h,k)}_{q}\in \{0,1\}^{|V_q|\times N_h}$ records the state of alignments for $R^{h,k}_{q}$, and satisfies
\begin{equation}
M^{h,k}_{q}(j,n)=\left\{
\begin{array}{cl}
1   & \mathrm{if} \ R^{h,k}_{q} (j,n) \ \mathrm{is \ the \ smallest } \\
    & \mathrm{element \ in \ row} \ n; \ \mathrm{and} \  |{\mathcal{S}}_{j}^k|\neq 0;\\
0   & \mathrm{otherwise}.
\end{array} \right.
\label{CoMatrix2}
\end{equation}
Based on Eq.(\ref{CoMatrix1}) and Eq.(\ref{CoMatrix2}), the $h$-level correspondence matrix $M^{(h,k)}_{p;q}\in \{0,1\}^{|V_p|\times |V_q|}$, that records the state of correspondence information between pairwise vertices of $G_p$ and $G_q$, is defined as
\begin{equation}
\mathcal{M}^{(h,k)}_{p;q}=(M^{h,k}_{p})( {M^{h,k}_{q}})^T.\label{tm}
\end{equation}
For the $h$-level correspondence matrix $\mathcal{M}^{(h,k)}_{p;q}$, the rows index the vertices of $G_p$, and the columns index the the vertices of $G_q$. If $\mathcal{M}^{(h,k)}_{p;q}(i,j)=1$, there is an one-to-one correspondence between the vertices $v_{i}\in V_p$ and $v_{j}\in V_q$, i.e., we say that they are aligned or matched. \hfill$\Box$

Note that, the vertex alignment information identified by $\mathcal{M}^{(h,k)}_{p;q}$ is transitive, i.e., for three vertices $u$, $v$ and $w$, if $u$ and $v$ are aligned, and $v$ and $w$ are aligned, then $u$ and $w$ are also aligned. This is because $M^{(h,k)}_{p;q}$ identifies the vertex correspondences by evaluating whether the vertices are aligned to the same set of $h$-level prototype representations $\mathbf{{PR}}^{h,k}\in \mathbb{{PR}}^{H,k}$. Finally, by hierarchically aligning each graph to the set of different $h$-level prototype representations $\mathbf{{PR}}^{h,k}$ from the $H$-hierarchical prototype representations $\mathbb{{PR}}^{H,k}$, we obtain a family of $H$-hierarchical transitive vertex correspondence matrices between $G_p$ and $G_q$ as
\begin{equation}
\mathbf{M}^{(H,k)}_{p;q}=\{\mathcal{M}^{(1,k)}_{p;q}, \cdots, {\mathcal{M}}^{(h,k)}_{p;q},\cdots,{\mathcal{M}}^{(H,k)}_{p;q}    \}.\label{HM}
\end{equation}

\noindent\textbf{Remarks:} The procedure of computing the family of $H$-hierarchical correspondence matrices $\mathbf{M}^{(H,k)}_{p;q}$ is \textbf{completely unsupervised}, since we do not utilize any class labels of the graphs in $\mathbf{G}$ during the computational process.

\subsection{The Hierarchical Transitive-Aligned Kernel}\label{KernelDefinition}
We develop a new Hierarchical Transitive-Aligned Kernel (HTAK) for graphs, based on the $H$-hierarchical transitive vertex correspondence matrices between graphs


\noindent\textbf{Definition (The HTAK kernel):} For the set of graphs $\mathbf{G}$, we commence by computing the $k$-dimensional DB representations of the vertices over all graphs in $\mathbf{G}$, as the $0$-level prototype representations $\mathbf{{PR}}^{0,k}$. Based on $\mathbf{{PR}}^{0,k}$ and the definition in Section~\ref{s2}, we generate a family of $H$-hierarchical prototype representations as $$\mathbb{{PR}}^{H,k}=\{\mathbf{{PR}}^{1,k}, \ldots,\mathbf{{PR}}^{h,k},\ldots, \mathbf{{PR}}^{H,k}\},$$ where $\mathbf{{PR}}^{h,k}$ represents the set of $h$-level prototype representations, and $1\leq h \leq H$. For a pair of graphs $G_p$ and $G_q$ from $\mathbf{G}$, by aligning the vertices of $G_p$ and $G_q$ to the sets of different $h$-level prototype representations $\mathbf{{PR}}^{h,k} \in \mathbb{{PR}}^{H,k}$, we compute the family of $H$-hierarchical transitive vertex correspondence matrices as $$\mathbf{M}^{(H,k)}_{p;q}=\{\mathcal{M}^{(1,k)}_{p;q}, \cdots, {\mathcal{M}}^{(h,k)}_{p;q},\cdots,{\mathcal{M}}^{(H,k)}_{p;q}\}$$ between $G_p(V_p,E_p)$ and $G_q(V_q,E_q)$ based on Eq.(\ref{HM}). With $\mathbf{M}^{(H,k)}_{p;q}$ to hand, the proposed HTAK kernel $k_{HTAK}^{(H)}$ between $G_p$ and $G_q$ is defined as
\begin{align}
k_{\mathrm{HTAK}}^{(H)}(G_p,G_q)=\sum_{h=1}^{H} \sum_{k=1}^{K}  \sum_{i=1}^{|V_p|} \sum_{j=1}^{|V_q|}
{\mathcal{M}}^{(h,k)}_{p;q}(i,j),\label{KernelF}
\end{align}
where $K$ is the greatest value of the parameter $k$ (i.e., $k$ varies from $1$ to $K$). As we have stated in Section~\ref{s2}, the parameter $k$ indicates the dimension of the vectorial vertex representations, and we propose to employ the $k$-dimensional DB representations of vertices as the vectorial vertex representations~\cite{DBLP:journals/pr/BaiH14}. Since the DB representations are computed based on the $\widetilde{k}$-layer expansion subgraphs ($1\leq \widetilde{k} \leq k$), the greatest value $K$ of the parameter $k$ corresponds to that of the longest shortest path between vertices over all graphs in $\mathbf{G}$. Eq.(\ref{KernelF}) indicates that $k_{\mathrm{HTAK}}^{(H)}(G_p,G_q)$ counts the number of aligned vertex pairs between $G_p$ and $G_q$ over all the $h$-level vertex correspondence matrices $\mathcal{M}^{(h,k)}_{p;q}\in \mathbf{M}^{(H,k)}_{p;q}$. \hfill$\Box$

\noindent\textbf{Lemma.} \emph{The kernel $k_{\mathrm{HTAK}}^{(H)}$ is positive definite (\textbf{pd}).}


\noindent\textbf{Proof.} Intuitively, the proposed HTAK kernel $k_{\mathrm{HTAK}}^{(H)}$ is \textbf{pd}, since it counts pairs of aligned vertices over the $H$ correspondence matrices $\mathcal{M}^{(h,k)}_{p;q}\in \mathbf{M}^{(H,k)}_{p;q}$ and the correspondence information identified by the proposed kernel is transitive. More formally, for the graph $G_p\in \mathbf{G}$, let ${F}^{(h,k)}(G_p)$ be a $N_h$-dimensional feature vector that counts the number of vertices aligned to the corresponding $h$-level prototype representations $\mathbf{{PR}}^{h,k}\in \mathbb{{PR}}^{H,k}$, and
\begin{align}
{F}^{(h,k)}(G_p)=&[\sum_{i=1}^{|V_p|} {M}^{h,k}_{p}(i,1), \ldots, \sum_{i=1}^{|V_p|} {M}^{h,k}_{p}(i,n),  \nonumber     \\
&\ldots,\sum_{i=1}^{|V_p|} {M}^{h,k}_{p}(i,N_h)]^T,\label{FVec1}
\end{align}
where the $n$-th element $\sum_{i=1}^{|V_p|} {M}^{h,k}_{p}(i,n)$ of ${F}^{(h,k)}(G_p)$ counts the number of vertices (from $G_p$) that are all aligned to the $n$-th $h$-level prototype representation ${\mathrm{PR}}^{h,k}_n \in \mathbf{{PR}}^{h,k}$, and ${M}^{h,k}_{p}(i,n)$ is defined by Eq.(\ref{CoMatrix1}). Similarly, for the graph $G_q$, we have the feature vector ${F}^{(h,k)}(G_q)$ as
\begin{align}
{F}^{(h,k)}(G_q)=&[\sum_{j=1}^{|V_q|} {M}^{h,k}_{q}(j,1), \ldots, \sum_{i=1}^{|V_q|} {M}^{h,k}_{q}(j,n),  \nonumber     \\
&\ldots,\sum_{i=1}^{|V_q|} {M}^{h,k}_{q}(j,N_h)]^T,\label{FVec2}
\end{align}
Based on Eq.(\ref{FVec1}) and Eq.(\ref{FVec2}), the HTAK kernel $k_{\mathrm{HTAK}}^{(H)}$ defined by Eq.(\ref{KernelF}) can be re-written as
\begin{align}
k_{\mathrm{HTAK}}^{(H)}(G_p,G_q)=\sum_{h=1}^{H} \sum_{k=1}^{K} \langle {F}^{(h,k)}(G_p),{F}^{(h,k)}(G_q)\rangle,\label{KernelDP}
\end{align}
where $\langle {F}^{(h,k)}(G_p),{F}^{(h,k)}(G_q)\rangle$ is an inner product, i.e., it is a \textbf{pd} linear kernel. As a result, the kernel $k_{\mathrm{HTAK}}^{(H)}$ can be seen as a kernel that sums the linear kernels $\langle {F}^{(h,k)}(G_p),{F}^{(h,k)}(G_q)\rangle$, and is thus \textbf{pd}. \hfill$\blacksquare$


\subsection{Discussions of the Proposed Kernel}
The new vertex alignment kernel $k_{\mathrm{HTAK}}^{(H)}$ has some important properties that are not available for some existing state-of-the-art graph kernels.

\textbf{First}, unlike the existing alignment kernels~\cite{DBLP:conf/icml/FrohlichWSZ05,DBLP:conf/icml/Bai0ZH15,DBLP:conf/ijcai/BaiZW0H15,DBLP:conf/iciap/BaiZR0H15,DBLP:journals/pr/NeuhausB06} that can also identify correspondence information between vertices or edges, the aligned vertices identified by the proposed HTAK kernel $k_{\mathrm{HTAK}}^{(H)}$ are transitive. This is because, as we have stated in Section \ref{GraphMatching}, the vertex alignment method employed in the proposed kernel can transitively align vertices between graphs. As a result, the proposed HTAK kernel $k_{\mathrm{HTAK}}^{(H)}$ not only overcomes the shortcoming of ignoring structural correspondences arising in most R-convolution kernels, but also reflects more precise correspondence information than the existing alignment or matching kernels~\cite{DBLP:conf/icml/FrohlichWSZ05,DBLP:conf/icml/Bai0ZH15,DBLP:conf/ijcai/BaiZW0H15,DBLP:conf/iciap/BaiZR0H15,DBLP:conf/sspr/BaiR0H14,DBLP:journals/pr/NeuhausB06}.

\textbf{Second}, as Fr{\"{o}}hlich et al.~\cite{DBLP:conf/icml/FrohlichWSZ05} have stated, the transitive alignment step is necessary to guarantee the positive definiteness of alignment kernels. Thus, the proposed HTAK kernel $k_{\mathrm{HTAK}}^{(H)}$ guarantees the positive definiteness that is not available to the aforementioned alignment kernels~\cite{DBLP:conf/icml/FrohlichWSZ05,DBLP:conf/icml/Bai0ZH15,DBLP:conf/ijcai/BaiZW0H15,DBLP:conf/iciap/BaiZR0H15,DBLP:conf/sspr/BaiR0H14,DBLP:journals/pr/NeuhausB06}.

\textbf{Third}, the computation of the proposed HTAK kernel $k_{\mathrm{HTAK}}^{(H)}$ for a pair of graphs incorporates the information over all graphs under comparisons. This is because $k_{\mathrm{HTAK}}^{(H)}$ is computed by hierarchically aligning the vertices of each graph to the different $h$-level prototype representations of the family of $H$-hierarchical prototype representations, that is hierarchically identified by $\kappa$-means method over the $k$-dimensional vectorial vertex representations over all graphs in $\mathbf{G}$, i.e., $k_{\mathrm{HTAK}}^{(H)}$ is not only computed though each individual pair of graphs. By contrast, most existing graph kernels only capture graph characteristics for each pair of graphs~\cite{DBLP:conf/icdm/BorgwardtK05,DBLP:conf/icml/Bach08,GarterCOLT2003,DBLP:journals/tnn/AzizWH13,shervashidze2010weisfeiler,DBLP:conf/icml/CostaG10}. As a result, the proposed kernel $k_{\mathrm{HTAK}}^{(H)}$ may reflect richer graph characteristics.

Finally, note that, since the basics of the proposed kernel $k_{\mathrm{HTAK}}^{(H)}$ is based on $k$-dimensional DB representations of vertices than do not encapsulate any vertex or edge label information. The proposed kernel $k_{\mathrm{HTAK}}^{(H)}$ cannot accommodate the vertex or edge label information. However, we can still perform $k_{\mathrm{HTAK}}^{(H)}$ on attributed graphs by focusing on topological information without vertex/edge labels.

\subsection{Computational Analysis}

For the set of $T$ graphs $\mathbf{G}$ each of which has $n$ vertices and $m$ edges, computing the proposed kernel $k_{\mathrm{HTAK}}^{(H)}$ requires time complexity $O(HIN_1Tn   +   HT^2N_1+HTN_1n   +  Tn\log n +Tmn)$, where $H$ corresponds to the set number of different $h$-level prototype representations from the family of $H$-hierarchical prototype representations, $I$ is the iteration number for the $\kappa$-means method, and $N_1$ is the number of $1$-level prototype representations in $\mathbf{{PR}}^{1,k}$. This is because computing the required $k$-dimensional DB representations of vertices (i.e., the $0$-level prototype representations $\mathbf{{PR}}^{0,k}$) relies on the shortest path computation on each graph, and thus requires time complexity $O(Tn\log n +Tmn)$. Computing the $N_h$ $h$-level prototype representations of $\mathbf{{PR}}^{h,k}$ relies on $\kappa$-means method on the last $N_{h-1}$ $h-1$-layer prototype representations of $\mathbf{{PR}}^{h-1,k}$. Since $N_1\gg N_2 \gg\cdots \gg N_H$, the whole process requires time complexity $O(HIN_1Tn)$. Calculating the kernel value between graphs relies on computing the $k$-level correspondence matrix $M^{(h,k)}_{p}\in \{0,1\}^{|V_p|\times N_h}$ in terms of each set of $h$-level prototype representations $\mathbf{{PR}}^{h,k}$, and counting the number of vertices of each graph aligning to the $N_h$ prototype representations in $\mathbf{{PR}}^{h,k}$. Since $N_1\gg N_2 \gg\cdots \gg N_H$, the whole process requires time complexity $O(HT^2N_1+HTN_1n )$.  As a result, the whole time complexity of computing the proposed kernel $k_{\mathrm{HTAK}}^{(H)}$ over all $T$ graphs of $\mathbf{G}$ requires time complexity $O(HIN_1Tn   +   HT^2N_1+HTN_1n   +  Tn\log n +Tmn)$. Note that, in this work, we employ the fastest $K$-means MATLAB implementation developed by Deng Cai~\cite{matlabKmeans}, and the default number of $I$ is $100$. Moreover, in this work, most graphs are sparse graphs (i.e., $n< m\ll n^2$) and $H$ is set as $5$. As a result, the whole time complexity is approximately $O(N_1Tn   +   T^2N_1+TN_1n   +Tn^2)$, indicating that our kernel can usually be computed in a polynomial time.

\begin{table*}
\centering {
\scriptsize
\caption{Information of the graph based computer vision (CV), bioinformatics (Bio), and social network (SN) datasets}\label{T:GraphInformation}

\begin{tabular}{|c||c||c||c||c||c||c||c||c||c|}

  \hline
 ~Datasets ~           &~BAR31  ~ & ~BSPHERE31~&  ~GEOD31~  & ~MUTAG~   & ~NCI1~        & ~CATH2~      & ~COLLAB  ~ & ~IMDB-B~   & ~IMDB-M~  \\ \hline \hline

  ~Max \# vertices~    &~ $220$~  & ~$227$~    &  ~$380$~   & ~$28$~    &  ~$111$~       &  ~$568$~    & ~$492$~    & ~$136$~     & ~$89$~ \\ \hline

  ~Mean \# vertices~   &~ $95.42$~& ~$99.83$~  &  ~$57.42$~ &  ~$17.93$~&  ~$29.87$~     &  ~$308.03$~ & ~$74.49$~  & ~$19.77$~   & ~$13.00$~      \\  \hline

  ~\# graphs~          &~ $300$~  & ~$300$~    &  ~$300$~   &  ~$188$~  &  ~$4110$~      &  ~$190$~    & ~$5000$~   &  ~$1000$~   & ~$1500$~     \\  \hline

  ~\# classes ~        &~ $15$~   & ~$15$~     &  ~$15$~    &  ~$2$~    &  ~$2$~         &  ~$2$~      &  ~$3$~     & ~$2$~       &  ~$3$~     \\  \hline

 ~Description~        &~ $\mathrm{CV}$~& ~$\mathrm{CV}$~ &  ~$\mathrm{CV}$~ &  ~$\mathrm{Bio}$~&  ~$\mathrm{Bio}$~ &  ~$\mathrm{Bio}$~      &  ~$\mathrm{SN}$~     & ~$\mathrm{SN}$~       &  ~$\mathrm{SN}$~     \\  \hline
\end{tabular}
}\vspace{-0pt}
\end{table*}

\section{Experiments}\label{s4}

We evaluate the proposed HTAK kernels on nine benchmark graph datasets from computer vision, bioinformatics, and social networks. These datasets include: BAR31, BSPHERE31, GEOD31, MUTAG, NCI1, CATH2, COLLAB, IMDB-B, and IMDB-M. Here the BAR31, BSPHERE31 and GEOD31 datasets are all abstracted from the SHREC 3D Shape database, that consists of $15$ classes and 20 individuals per class~\cite{DBLP:conf/dgci/BiasottiMMPSF03}. Specifically, we establish the BAR31, BSPHERE31 and GEOD31 datasets through three mapping functions,i.e., a) ERG barycenter: distance from the center of mass/barycenter, b) ERG bsphere: distance from the center of the sphere that circumscribes the object, and c) ERG integral geodesic: the average of the geodesic distances to the all other points. On the other hand, other datasets are all available on the website http://graphkernels.cs.tu-dortmund. More details of these datasets are shown in Table.\ref{T:GraphInformation}.

\subsection{Experiments on Graph Classification}

\noindent\textbf{Experimental Setup:} We evaluate the performance of the proposed HTAK kernel in terms of graph classification
problems on the aforementioned nine benchmark graph datasets. We also compare our kernel with a) five alternative state-of-the-art graph kernels and b) four alternative state-of-the-art deep learning methods for graphs. Specifically, \textbf{the graph kernels include} 1) the aligned subtree kernel (ASK)~\cite{DBLP:conf/icml/Bai0ZH15}, 2) the Weisfeiler-Lehman subtree kernel (WLSK)~\cite{DBLP:journals/jmlr/ShervashidzeSLMB11}, 3) the shortest path graph kernel (SPGK)~\cite{DBLP:conf/icdm/BorgwardtK05}, 4) the graphlet count graph kernel~\cite{DBLP:journals/jmlr/ShervashidzeVPMB09} with graphlet of size $4$ (GCGK), and 5) the Jensen-Tsallis q-difference kernel (JTQK) \cite{DBLP:conf/pkdd/Bai0BH14} with $q=2$. On the other hand, \textbf{the deep learning methods include} 1) the deep graph convolutional neural network (DGCNN)~\cite{DBLP:conf/aaai/ZhangCNC18}, 2) the PATCHY-SAN based convolutional neural network for graphs (PSGCNN)~\cite{DBLP:conf/icml/NiepertAK16}, 3) the diffusion convolutional neural network (DCNN)~\cite{DBLP:conf/nips/AtwoodT16}, and 4) the deep graphlet kernel (DGK)~\cite{DBLP:conf/kdd/YanardagV15}.

For the WLSK kernel and the JTQK kernel, we set the highest dimension (i.e., the highest height of subtrees) of the Weisfeiler-Lehman isomorphism (for the WLSK kernel) and the tree-index method (for the JTQK kernel) as $10$, based on the statements of the authors in~\cite{DBLP:conf/pkdd/Bai0BH14,DBLP:journals/jmlr/ShervashidzeSLMB11}. For the ASK kernel, we set the highest layer of the required DB representation as $50$ based on~\cite{DBLP:conf/icml/Bai0ZH15}, to guarantee the best performance. For each kernel, we compute the kernel matrix on each graph dataset. We perform a $10$-fold cross-validation where the classification accuracy is computed using a C-Support Vector Machine (C-SVM). In particular, we make use of the LIBSVM library\cite{ChangLinSVM2001}. For each dataset and each kernel, we compute the optimal C-SVMs parameters. We repeat the whole experiment 10 times and report the average classification accuracy ($\pm$ standard error) in Table~\ref{T:Classification}. Note that, for the proposed HTAK kernel we vary the parameter $H$ from $1$ to $5$. Thus, for each dataset we compute $5$ kernel matrices for the HTAK kernel. The classification accuracy for each time is thus the average accuracy over the $5$ kernel matrices. Moreover, for the proposed HTAK kernel on each dataset, we set the parameter $N_h$ as $N_h=0.2  N_{h-1}$, where $h$ varies from $1$ to $5$ and $N_0$ corresponds to the vertex number over all graphs in the dataset.


For the alternative deep learning methods, we report the best results for the DGCNN, PSGCNN, DCNN, DGK models from their original papers. Moreover, note that the PSGCNN model can leverage additional edge features, most of the graph datasets and the alternative methods do not leverage edge features. Thus, we do not report the results associated with edge features in the evaluation. The classification accuracies and standard errors for each deep learning method are shown in Table.\ref{T:ClassificationDL}. Finally, note that, as we have stated in Section~\ref{KernelDefinition}, the computation of the HTAK kernel for a pair of graphs incorporates the information over all graphs under comparisons. Thus the proposed HTAK kernel can also incorporate the test graphs into the training process of C-SVMs. In this sense, the proposed HTAK kernel can be seen as an instance of \textbf{transductive learning}~\cite{gammerman1998learning} (i.e., we transductively train the C-SVM), where all the graphs available (both from the training and test sets) are used to compute the graph centroid representations. However, note that we \textbf{do not} observe the class labels of the test graphs during the training. Finally, note that, some methods are not evaluated by the original authors on some datasets, thus we do not exhibit these results.


\begin{table*}
\vspace{-0pt}
\centering {
\tiny
\caption{Classification Accuracy (In $\%$ $\pm$ Standard Error) for Comparisons with Graph Kernels.}\label{T:Classification}
\vspace{0pt}
\begin{tabular}{|c||c||c||c||c||c||c||c||c||c|}

  \hline
 ~Datasets ~&~BAR31  ~       & ~BSPHERE31~    &~GEOD31~       & ~MUTAG~        & ~NCI1~            & ~CATH2~              & ~COLLAB  ~ & ~IMDB-B~   & ~IMDB-M~           \\ \hline \hline

  ~\textbf{HTAK}~   &~ $71.00\pm.45$~& ~$\textbf{62.90}\pm.65$~&~$\textbf{47.80}\pm.49$~&~$87.32\pm.60$~ &  ~$79.01\pm.14$~   &  ~$\textbf{87.89}\pm.71$~  &~$\textbf{79.87}\pm0.15$~ &~$\textbf{72.89}\pm0.56$~ &  ~$50.23\pm0.18$~   \\ \hline

  ~ASK ~    &~ $\textbf{73.10}\pm.67$~& ~$60.30\pm.44$~&~$46.21\pm.69$~&~$\textbf{87.50}\pm.65$~  &  ~$78.47\pm.12$~ &  ~$78.52\pm.67$~  &~$77.53\pm0.31$~ &~$70.38\pm0.72$~ &  ~$50.12\pm0.51$~   \\  \hline

  ~WLSK~    &~ $58.53\pm.53$~& ~$42.10\pm.68$~&~$38.20\pm.68$~&~$82.88\pm0.57$~ &  ~$84.77\pm.13$~ &  ~$67.36\pm.63$~  &~$77.39\pm0.35$~ &~$71.88\pm0.77$~ &  ~$49.50\pm0.49$~    \\  \hline

  ~SPGK~    &~ $55.73\pm.44$~& ~$48.20\pm.76$~&~$38.40\pm.65$~&~$83.38\pm0.81$~  &  ~$74.21\pm.30$~&  ~$81.89\pm.63$~  &~$58.80\pm0.2$~  &~$71.26\pm1.04$~ &  ~$\textbf{51.33}\pm0.57$~  \\  \hline

  ~GCGK ~   &~ $23.40\pm.60$~& ~$18.80\pm.50$~&~$22.36\pm.55$~&~$82.04\pm.39$~  &  ~$63.72\pm.12$~ &  ~$73.68\pm1.09$~  &~$-$~  &~$-$~ &~$-$~  \\  \hline

  ~JTQK ~   &~ $60.56\pm.35$~& ~$46.93\pm.61$~&~$40.10\pm.46$~&~$85.50\pm.55$~  &  ~$\textbf{85.32}\pm.14$~ &  ~$68.70\pm.69$~  &~$76.85\pm0.40$~ &~$72.45\pm0.81$~ &  ~$50.33\pm0.49$  \\  \hline




\end{tabular}
} \vspace{-0pt}
\end{table*}

\begin{table*}
\vspace{-0pt}
\centering {
\tiny
\caption{Classification Accuracy (In $\%$ $\pm$ Standard Error) for Comparisons with Deep Learning Methods.}\label{T:ClassificationDL}
\vspace{0pt}
\begin{tabular}{|c||c||c||c||c||c|}

  \hline
 ~Datasets~& ~MUTAG~        & ~NCI1~              & ~COLLAB  ~      & ~IMDB-B~        & ~IMDB-M~           \\ \hline \hline

  ~\textbf{HTAK}~   &~$87.32\pm.60$~ &  ~$\textbf{79.01}\pm.14$~    &~$\textbf{79.87}\pm0.15$~ &~$\textbf{72.89}\pm0.56$~ &  ~$\textbf{50.23}\pm0.18$~   \\ \hline

  ~DGCNN~  &~$85.83\pm1.66$~&  ~$74.44\pm.47$~    &~$73.76\pm0.49$ ~& ~$70.03\pm0.86$ & ~$47.83\pm0.85$~   \\  \hline

  ~PSGCNN~ &~$\textbf{88.95}\pm4.37$~&  ~$76.34\pm1.68$~    & ~$72.60\pm2.15$ ~& ~$71.00\pm2.29$ & ~$45.23\pm2.84$ ~    \\  \hline

  ~DCNN~   &~$66.98$~       &  ~$56.61\pm1.04$~    & ~$52.11\pm0.71$ ~& ~$49.06\pm1.37$ & ~$33.49\pm1.42$ ~ \\  \hline

  ~DGK ~   &~$82.66\pm1.45$~ &  ~$62.48\pm.25$~    & ~$73.09\pm0.25$~& ~$66.96\pm0.56$ & ~$44.55\pm0.52$ ~\\  \hline


\end{tabular}
} \vspace{-10pt}
\end{table*}
\noindent\textbf{Results and Discussions:} In terms of the classification accuracy, we observe that our HTAK kernel can outperform the alternative graph kernels and deep learning methods on most datasets. For the alternative graph kernel methods, only the accuracies of the ASK kernel on the BAR31 and MUTAG datasets, and the accuracy of the SPGK kernel on the IMDB-M dataset as well as that of the JTQK kernel on the NCI1 dataset are higher than the proposed HTAK kernel. On the other hand, for the alternative deep learning methods, only the PSGCNN model on the MUTAG dataset is higher than the proposed HTAK kernel.

In fact, the WLSK, ASK and JTQK kernels, as well as the alternative deep learning approaches can all accommodate the vertex label information, i.e., they can accommodate attributed graphs. By contrast, the proposed HTAK kernel is designed for un-attributed graphs and can cannot associate with any vertex label information. On the other hand, only these deep learning methods can provide an end-to-end learning framework for graph classification. By contrast, the proposed HTAK kernel associated with the C-SVM can only provide a shallow learning framework. However, even under such disadvantageous situations, the proposed HTAK kernel can still outperform these methods on most datasets. This indicate that the proposed kernel can learn better topological characteristics of graphs than the remaining alternative methods, through the family of $H$-hierarchical prototype representations that represent a reliable pyramid of the original vertex representations over all graphs at different levels (i.e., the prototype representations of different $h$-levels).

The reasons for the effectiveness are fourfold. \textbf{First}, unlike the alternative WLSK, SPGK, GCGK and JTQK kernels that ignore the correspondences information between substructures, the proposed HTAK kernel can hierarchically identify the vertex correspondence information through the $H$-hierarchical prototype representations. \textbf{Second}, compared to the ASK kernel, the correspondence information identified by the HTAK kernel are transitive. By contrast, the ASK kernel cannot guarantee the transitivity.  As a result, the HTAK kernel can capture more precise information for graphs than the ASK kernel. \textbf{Third}, unlike alternative kernels, only the proposed kernel incorporates the information of all graphs under comparisons into the kernel computation. The HTAK kernel thus reflects richer graph characteristics. \textbf{Fourth}, similar to the WLSK, SPGK, GCGK and JTQK kernels, all the alternative deep learning methods also do not associate with the structural correspondence information into the learning framework. Overall, the above observations demonstrate the effectiveness of the proposed HTAK kernel.

\section{Conclusions}\label{s5}

In this paper, we develop a new Hierarchical Transitive-Aligned kernel for graphs, that can transitively align the vertices between graphs through a family of $H$-hierarchical prototype graphs. Unlike most state-of-the-art graph kernels, this kernel not only overcomes the shortcoming of ignoring correspondence information between graphs, but also guarantees the transitivity between the correspondence information. Experimental evaluations have demonstrated the effectiveness of the proposed new transitive aligned kernel. The proposed kernel can outperform state-of-the-art graph kernels as well as the deep learning methods in terms of graph classifications.

Our future work is to further extend the proposed kernel for attributed graphs, so that the proposed kernel can accommodate the vertex label information into the computation, improving the performance the proposed kernel.

\section*{Acknowledgments}
This work is supported by the National Natural Science Foundation of China (Grant no. 61976235 and 61602535), the Open Projects Program of the National Laboratory of Pattern Recognition (NLPR), and the program for innovation research in Central University of Finance and Economics and the Youth Talent Development Support Program by Central University of Finance and Economics, No. QYP1908.

\balance


\bibliographystyle{IEEEtran}
\bibliography{example_paper}

\end{document}